\documentclass[twocolumn]{aastex631}
\graphicspath{{./},{./figures/}}
\usepackage{amsmath}
\usepackage{mathptmx,txfonts}
\usepackage{booktabs}
\usepackage{bm}
\usepackage[capitalize]{cleveref}

\newcommand{\numax}{{\ensuremath{\nu_{\text{max}}}}}

\renewcommand{\edit}[2]{{\ifnum#1<3%
#2%
\else%
\textbf{#2}%
\fi}}

\begin{document}

\correspondingauthor{Christopher Lindsay}
\email{christopher.lindsay@yale.edu}

\title{Near-Core Acoustic Glitches are Not Oscillatory:\\ Consequences for Asteroseismic Probes of Convective Boundary Mixing}

\author[0000-0001-8722-1436]{Christopher J. Lindsay}
\affiliation{Department of Astronomy, Yale University, PO Box 208101, New Haven, CT 06520-8101, USA}
\author[0000-0001-7664-648X]{J. M. Joel Ong}
\affiliation{Institute for Astronomy, University of Hawai`i, 2680 Woodlawn Drive, Honolulu, HI 96822, USA}
\affiliation{Hubble Fellow}
\author[0000-0002-6163-3472]{Sarbani Basu}
\affiliation{Department of Astronomy, Yale University, PO Box 208101, New Haven, CT 06520-8101, USA}

\shortauthors{Lindsay, Ong \& Basu}
\shorttitle{Non-JWKB Treatment of Near-Core Glitches}

\begin{abstract}
 Asteroseismology has been used extensively in recent years to study the interior structure and physical processes of main sequence stars. We consider prospects for using pressure modes (p-modes) near the frequency of maximum oscillation power to probe the structure of the near-core layers of main sequence stars with convective cores by constructing stellar model tracks. Within our mass range of interest, the inner turning point of p modes as determined by the JWKB approximation evolves in two distinct phases during the main sequence, implying a sudden loss of near-core sensitivity during the discontinuous transition between the two phases. However, we also employ non-JWKB asymptotic analysis to derive a contrasting set of expressions for the effects that these structural properties will have on the mode frequencies, which do not encode any such transition. We show analytically that a sufficiently near-core perturbation to the stellar structure results in non-oscillatory, degree-dependent perturbations to the star's oscillation mode frequencies, contrasting with the case of an outer glitch. We also demonstrate numerically that these near-core acoustic glitches exhibit strong angular degree dependence, even at low degree, agreeing with the non-JWKB analysis, rather than the degree-independent oscillations which emerge from JWKB analyses. These properties have important implications for using p-modes to study near-core mixing processes for intermediate-mass stars on the main sequence, as well as for the interpretation of near-center acoustic glitches in other astrophysical configurations, such as red giants.
\end{abstract}


\section{Introduction} \label{sec:intro}

The long temporal baselines of the \textit{Kepler} \citep{Kepler_inst} and TESS \citep{TESS_inst} missions make the interiors of thousands of stars amenable to examination through the asteroseismology of individual mode frequencies in their photometric power spectra. Stars with convective envelopes, such as our Sun, oscillate in multiple modes excited by convective motions \citep{Goldreich1977a, Goldreich1977b}.\edit1{ The frequencies of these oscillation modes can trace stellar structure in the deep interior, thereby encoding information about the star's evolutionary state \citep[][and references therein]{Chaplin2013, Garcia2019}}. The internal modes of solar-type oscillators are generally classified as either p-modes --- where the restoring force is pressure --- or g-modes --- where the restoring force is gravity. Asymptotic analysis of wave propagation in stars under the Jeffreys-Wentzel-Kramers-Brillouin (JWKB) approximation \citep[see][]{Gough2007} indicates that all non-radial modes are limited in their sensitivity to different regions of the star, depending on their character. In Sun-like stars, p-modes and g-modes occur in different regions: p-modes propagate in the outer convective envelope, and g-modes in the core. The loci of these different classes of modes are governed by two characteristic frequencies: the Lamb frequency
\begin{equation}
    S_{\ell}^2 =  \frac{\ell(\ell+1)c_s^2}{r^2},
\end{equation}
 and the Brunt–Väisälä (or buoyancy) frequency
 \begin{equation}
     N^2 = -g\left(\frac{1}{\Gamma_1 P}\frac{dP}{dr}-\frac{1}{\rho}\frac{d\rho}{dr}\right);
 \end{equation}
 where $\ell$ is the angular degree of the mode, $c_s$ the sound speed, $P$ the pressure, $\rho$ the density, and $r$ is the radial coordinate. Waves which are higher in frequency than both these frequencies are p-modes (shaded in orange in \cref{fig:propagation}), while those that are lower in frequency than both are g-modes (shaded in blue). For stars on the main sequence, these p- and g-mode cavities are well separated both in frequency and spatially, so any normal modes with observable amplitudes (that is, with frequencies near that of maximum oscillation power, $\numax$) are purely acoustic (p-modes). The depth to which p-modes sample the stellar structure is set by the Lamb frequency of the corresponding $\ell$, which, due to its dependence on the sound speed, will also depend on the mean-molecular-weight gradient $\nabla_\mu$. Depending on the properties of near-core mixing, as well as how evolved the star is along the main sequence, the observed p-modes may therefore not penetrate deeply enough to reach the convective cores of main sequence stars, thereby, in principle, limiting the applicability of these p-modes for diagnosing the nature of such near-core mixing, under the WKB approximation. 
 
The interior locations of convective boundaries and abundant element ionization zones are frequently studied through asteroseismic "glitch" analysis. Steep variations in the \edit1{first adiabatic index, $\Gamma_1$, or in the} sound speed, are known to introduce an oscillatory component  ($\delta \nu$) to the frequencies of low angular degree stellar oscillation modes \citep[e.g.][]{Gough1988, Vorontsov1988, Gough1990, Basu1994}. \edit1{A number of investigations have explored the theoretical seismic consequences arising from sharp variations in the internal stellar structure at the convective envelope boundary \citep[e.g.,][]{Monteiro2000} or in the region of helium ionization \citep{Monteiro1998, Houdek2007}}. These glitch signals \edit1{are present across a range of stellar evolutionary states, and} have been used to investigate the \edit1{properties} of convection and ionization zones in both red giant stars \citep{Miglio2010, Corsaro2015, Vrard2015, Dreau2020} and Sun-like main-sequence stars \citep{Mazumdar2012, Mazumdar2014}. Since these \edit1{convective envelope and helium ionization zone} features are localized far outside the convective core, we will refer to these glitches in this work as "outer glitches". 
 
 \edit1{The sharp variations in stellar structure present at the boundaries of convective cores are also expended to leave a signature on the oscillation frequencies of a star. However, any such signature from a convective core will have different properties than those generated from an outer glitch, owing to their localization close to the stellar center, rather than surface. \citet{RoxburghVorontsov2001} investigated the expected seismic signatures resulting from a glitch in the neighborhood of a convective core, provided that the structural variation was located well within the mode propagation cavity, far from the turning point of the oscillations. Similar analyses were performed by \citet{Provost1993} in the case of structural variations at the boundary of Jupiter's core and by \citet{Audard1995} in the case of intermediate mass (1.7 - 2.0 $M_{\odot}$) stars with g-mode pulsations. In the case of a lower mass (1.2-1.5 $M_{\odot}$) main-sequence star though, the aforementioned works do not apply as the convective core is small, with its boundary located very close to the inner turning point of the oscillation modes.}

\edit1{\citet{Mazumdar2006} studied the seismic effects of small convective cores in stellar models and proposed a combination of small frequency separations with the goal of determining the presence of convective overshooting. A similar investigation was carried out by \citet{CunhaMetcalfe2007}, who found that the seismic signatures of small convective cores are non-oscillatory and frequency-dependent. They suggest a combination of frequency separation ratios that may have diagnostic potential for studying convective cores in real stars with high quality asteroseismic data. However, as with \citet{Mazumdar2006}, their proposed diagnostic combined information from modes of different degrees. As such, they were unable to investigate the angular degree dependence of the seismic signal (instead assuming a priori that it would only affect the radial modes). \citet{Brandao2010} further investigated these diagnostics to look for age-dependence. \citet{Cunha2011} built on the work of \citet{CunhaMetcalfe2007} and further investigated the seismic signatures of small convective cores. In particular, the work modelled the structural variation at the edge of convective cores in a more physically-motivated fashion to study the evolution of their seismic diagnostic as a star advances in age.}

\edit1{In this work, }we investigate \edit1{the near-core} locations available for study through low angular degree mode glitch signature analysis, both within \edit1{(section \ref{sec:methods})} and outside \edit1{(section \ref{sec:nonwkb})} the WKB approximation, using evolutionary tracks of stellar models. \edit1{We discuss our results and compare our work to previous studies of the seismic signatures of convective cores in section \ref{sec:discussion}.}

\section{WKB Analysis with Stellar Models} \label{sec:methods}

\begin{figure*}[htbp]
    \centering
    \includegraphics[width = 0.9\textwidth]{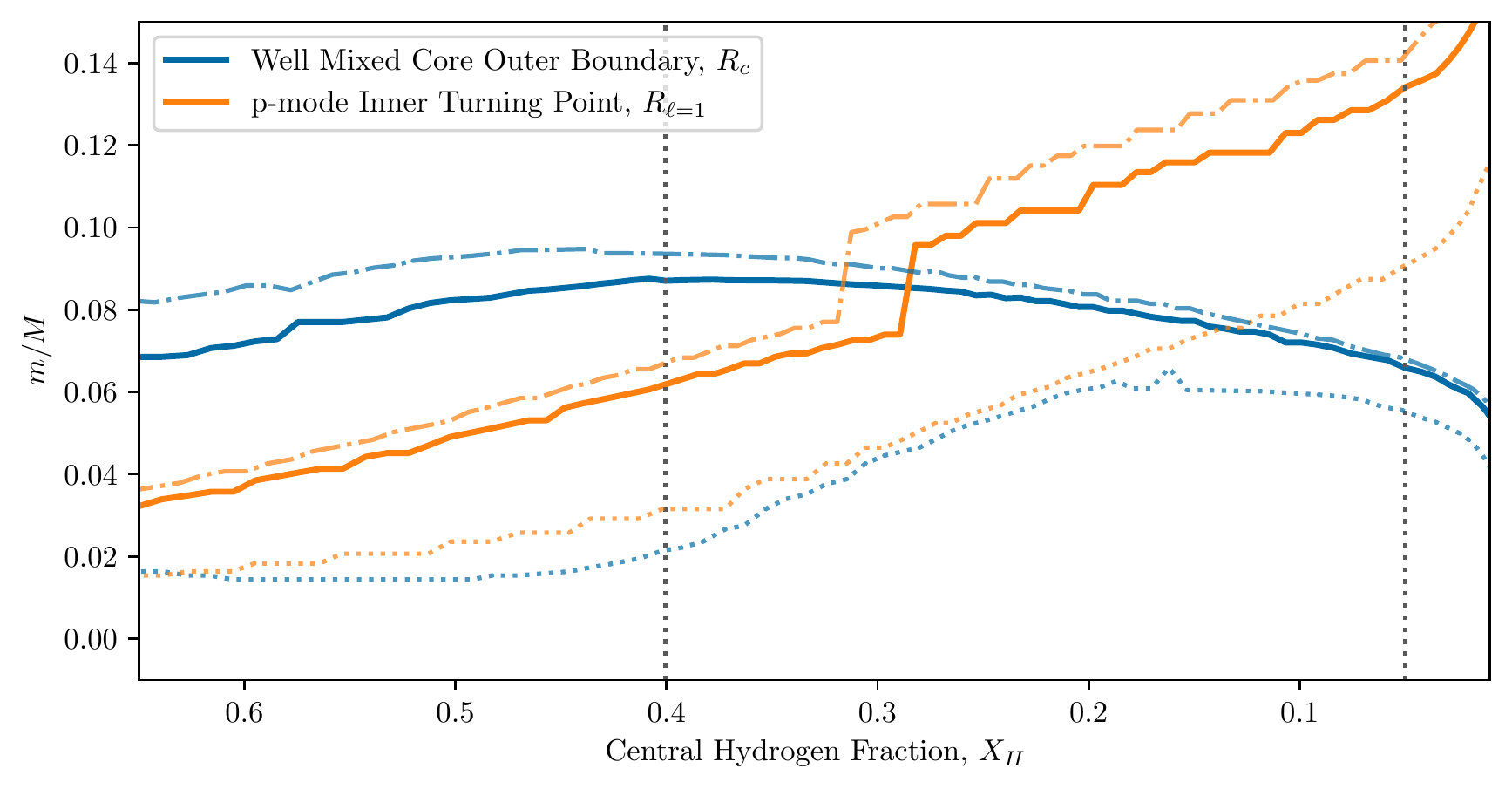}
    \caption{Evolution of the well-mixed core outer boundary (blue)  inner turning point (orange) of $\ell=1$ modes near $\nu_{\text{max}}$ in mass coordinates for 1.2 $M_{\odot}$ (dotted), 1.4 $M_{\odot}$ (solid), and 1.5 $M_{\odot}$ (dot dashed) mass model tracks. Evolution goes from left to right as central hydrogen fraction decreases along the main sequence. }
    \label{fig:figure}
\end{figure*}

To illustrate the evolution of the well-mixed core and p-mode penetration depths, we construct stellar model tracks with masses between 1.2 and 1.5 $M_{\odot}$ using MESA version r12778 \citep{Paxton2011, Paxton2013, Paxton2015, Paxton2018, Paxton2019}. We construct models using an Eddington-gray atmospheric boundary condition and the mixing-length prescription of \citet{CoxGiuli1968}. Elemental diffusion following the formulation of \citet{Thoul1994} was included with mass-dependent scaling \citep[see][]{Viani_2018}. We show results for 3 model tracks with $M = $ 1.2, 1.4, and 1.5 $M_{\odot}$, calculated with solar-calibrated initial values of Helium abundance ($Y_0 = 0.273$), metallicity \citep[relative to][]{GS98}, and mixing length ($\alpha_{\text{mlt}} = 1.81719$). MESA's implementation of overmixing \citep[cf. §2 of][]{Lindsay2022} from the convective core was also used, with $f_{ov} = 0.05$.

Within the WKB approximation, non-radial p-modes are assumed to only be sensitive to stellar structure within a mode cavity bounded on the inside by the WKB inner turning point, where the mode angular frequencies are equal to the Lamb frequency. For non-radial modes of the same frequency, this inner turning point's radius value increases with $\ell$, and is thus deepest for dipole modes. Accordingly, we show in \cref{fig:figure} the evolution of both the outer boundary of the well-mixed core, $R_c$, as well as of the inner turning points of dipole p-modes at $\numax$, $R_{\ell = 1}$, over the course of evolution along these tracks (parameterized by the central hydrogen fraction $X_H$). We define $R_c$ as the location where the chemical gradient $\nabla_\mu$ changes by more than 0.1 between adjacent mesh points, while $R_{\ell = 1}$ is the innermost point where $S_{\ell = 1} = 2 \pi \numax$. Locations are indicated with respect to the relative mass coordinate $m(r)/M$. 


From \autoref{fig:figure}, we see that for the 1.4 and 1.5 $M_{\odot}$ evolutionary tracks (solid and dot dashed lines), $R_{\ell = 1}$ begins increasing steadily with  evolution along the main sequence. This steady rise is interrupted by a sharp discontinuity just after reaching $X_H = 0.3$ for the 1.4 $M_{\odot}$ track and just before reaching $X_H = 0.3$ for the 1.5 $M_{\odot}$ track. After this jump in $R_{\ell = 1}$, the dipole p-mode inner turning point at $\nu_{\text{max}}$ lies outside the near-core layers of these stars, rendering them insensitive to this region. Conversely, this discontinuous jump in $R_{\ell=1}$ does not occur in the 1.2 $M_{\odot}$ track since, given our specific combination of model input parameters, the position of the p-mode inner turning point starts (at zero age main sequence) at approximately the same location as the well-mixed core outer boundary ($R_{\ell = 1} \approx R_c$). Therefore, for this example 1.2 $M_{\odot}$ track, non-radial modes may not be used (under the JWKB approximation) to probe the near-core layers no matter how far along the main sequence the star has evolved. 

\begin{figure*}[htbp]
    \centering
    \includegraphics[width = 0.9\textwidth]{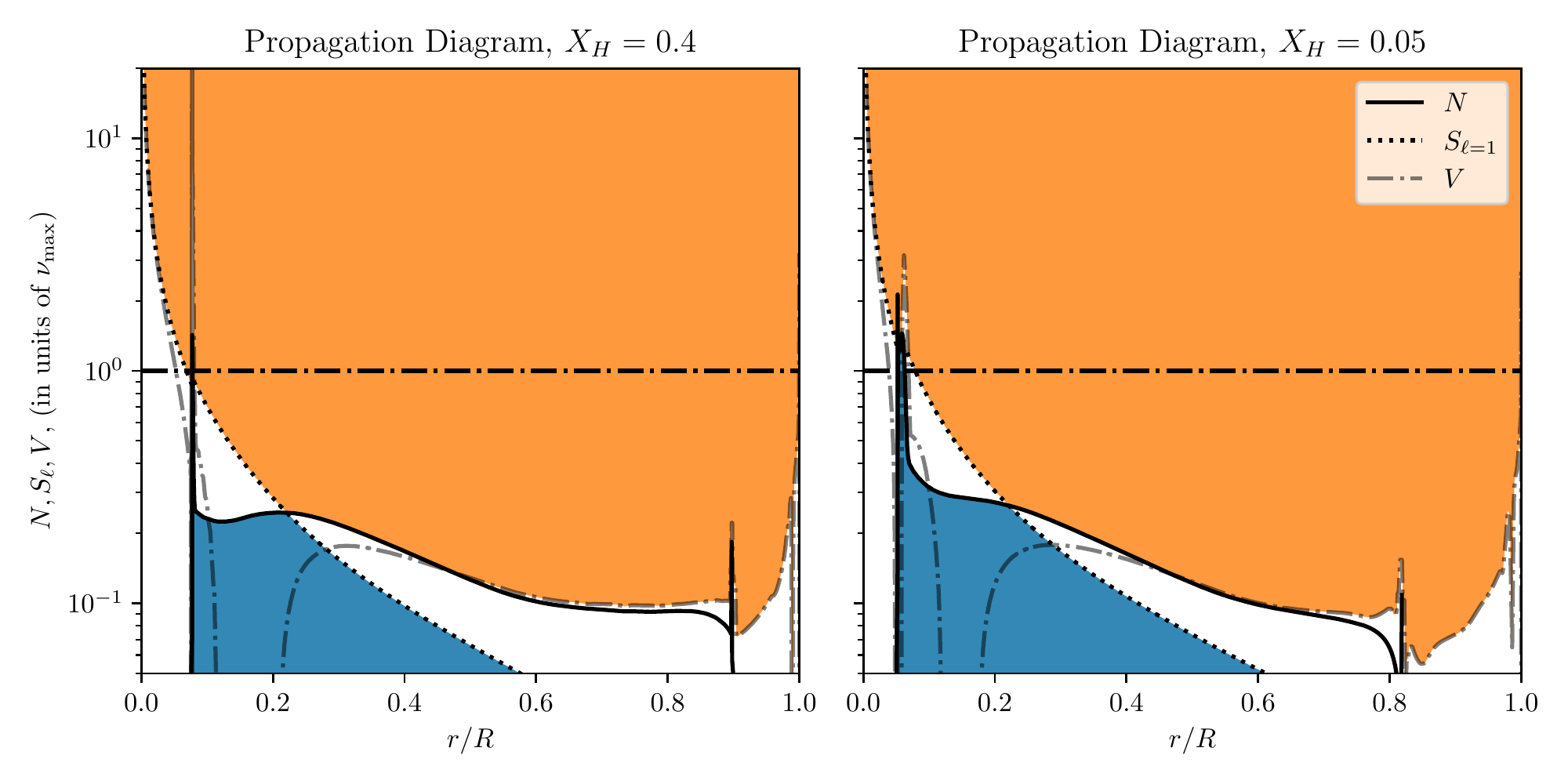}
    \includegraphics[width = 0.9\textwidth]{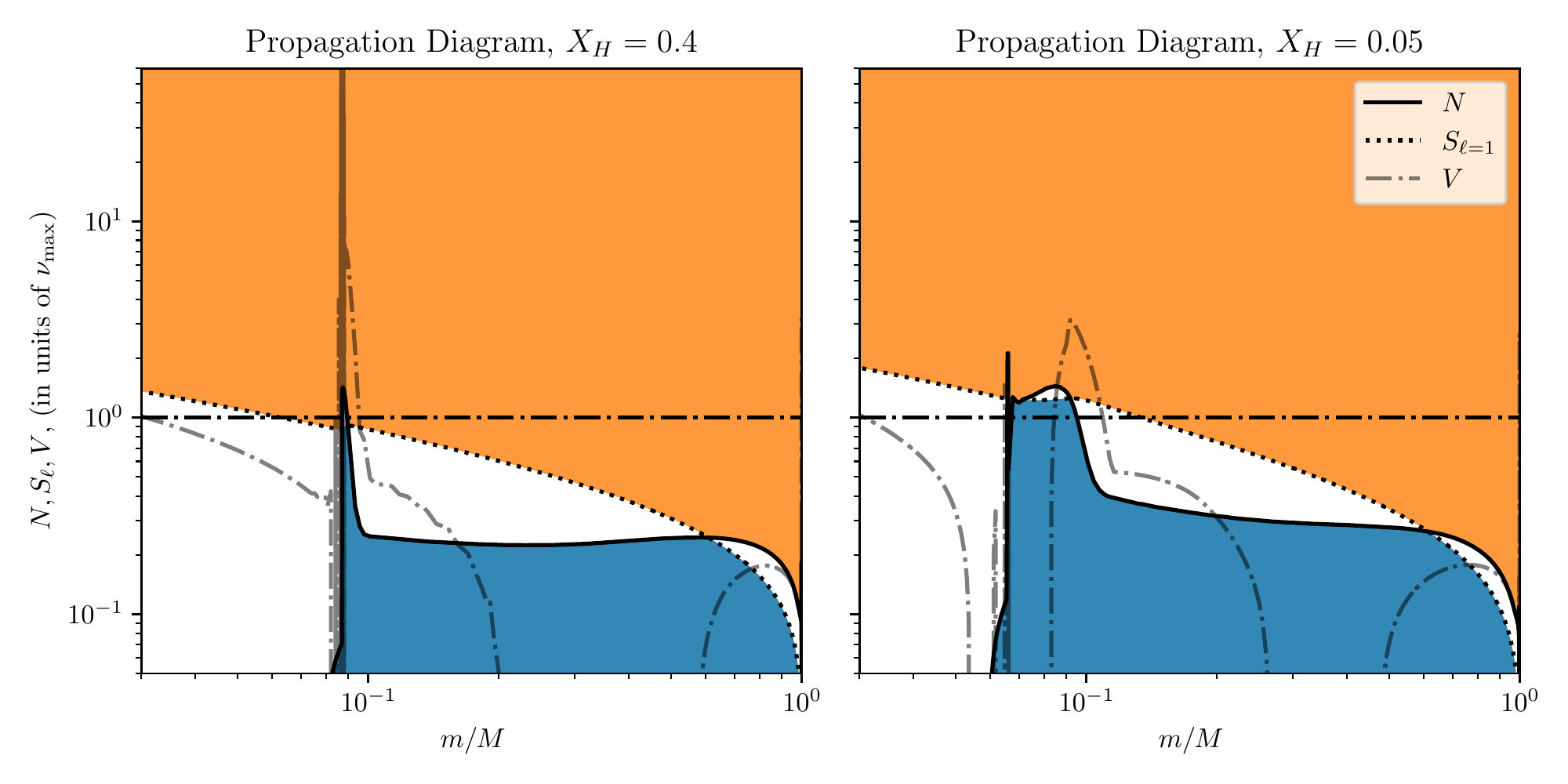}
    \caption{ \textbf{Upper Left Panel}: Propagation diagram for a 1.4 $M_{\odot}$ model with $X_H$ = 0.4 (left-most vertical gray dotted line in \cref{fig:figure}) showing, in units of $\nu_{\text{max}}$, the buoyancy frequency ($N$), the Lamb frequency ($S_{\ell = 1}$), and the acoustic potential, $V$, which is a characteristic frequency describing the propagation of a star's radial modes. The orange region of the propagation diagrams represent the regions where p-modes can propagate, while the blue regions represent the g-mode propagation regions. \edit2{In the outer layers of the star, the minimum frequency at which p-modes can propagate is governed by the critical frequency ($\nu_{\text{crit}}$ mirrors $V$ near the model's surface) in the outer layers.} \textbf{Upper Right Panel}: Same as the left panel but for the 1.4 $M_{\odot}$ model later in evolution (right gray dotted line in \cref{fig:figure}, $X_H = 0.05$)
    \textbf{Lower Panels}: Propagation diagrams for the same two models as in the upper panels, but the x-axis shows the relative mass coordinate in log-scale in order to show the near-core features in more detail. The acoustic potential, $V$, shows a sharp, localized peak at the position of the near-core glitch.}
    \label{fig:propagation}
\end{figure*}

These discontinuities in the evolution of the p-mode inner turning point ($R_{\ell = 1}$) emerge from kinks in the Lamb frequency profile, caused by the change in mean molecular weight gradient at the boundary of the star's convective core. To examine the underlying mechanism for this, we show propagation diagrams from the 1.4 $M_{\odot}$ track, before and after this discontinuous jump in $R_{\ell = 1}$, in \autoref{fig:propagation}. The spikes in buoyancy frequency ($N$, solid black lines), which correspond to local enhancements of $\nabla_\mu$, coincide with localized kinks in Lamb frequency ($S_{\ell=1}$, dotted lines). As the star evolves, these kinks move inwards, and the Lamb frequency at their location increases relative to $\nu_{\text{max}}$ (horizontal dot dashed line). When these kinks coincide with $\numax$, this results in a temporally discontinuous increase in $R_{\ell = 1}$. Since the pulsation wavefunction is assumed to decay exponentially in the WKB-evanescent region, this corresponds to a discontinuous reduction in the probing power of dipole modes to these near-core features on either side of this evolutionary boundary.

Unlike the non-radial modes, the radial ($\ell = 0$) modes are known to penetrate more deeply into the stellar interior. These modes admit description by an equation of Schr\"odinger form \citep[i.e. the ``normal form'' of][for JWKB analysis]{Gough2007}, with respect to the acoustic radial coordinate $t(r) = \int_0^r \mathrm d r / c_s$, where the acoustic potential function $V$ \textbf{(shown in \autoref{fig:propagation})}, is set by the stellar structure and determines the behavior of their wavefunctions near the center \citep[see e.g.][for a thorough discussion of the radial-mode acoustic potential]{Gough1993, Roxburgh2010, Ong2019}. Localized enhancements in this potential function are known to yield oscillatory signatures \cite[e.g.][]{Houdek2006}, known colloquially as ``glitches". Accordingly, we show this acoustic potential function in both propagation diagrams of \cref{fig:propagation} (scaled by $\numax$, using the gray dot-dashed lines). Sharply localized peak-like features in $V$ can be seen to emerge, corresponding to the locations in the star where chemical abundances vary rapidly with depth (near $m/M = 0.09$ and $m/M = 0.07$ for the $X_H = 0.4$ and $X_H = 0.05$ \autoref{fig:propagation} propagation diagrams, respectively). As such, these features must also have a direct effect on the radial-mode frequencies.

\section{Beyond the WKB Approximation} \label{sec:nonwkb}

\begin{figure*}[htbp]
    \centering
    \includegraphics[width = 0.9\textwidth]{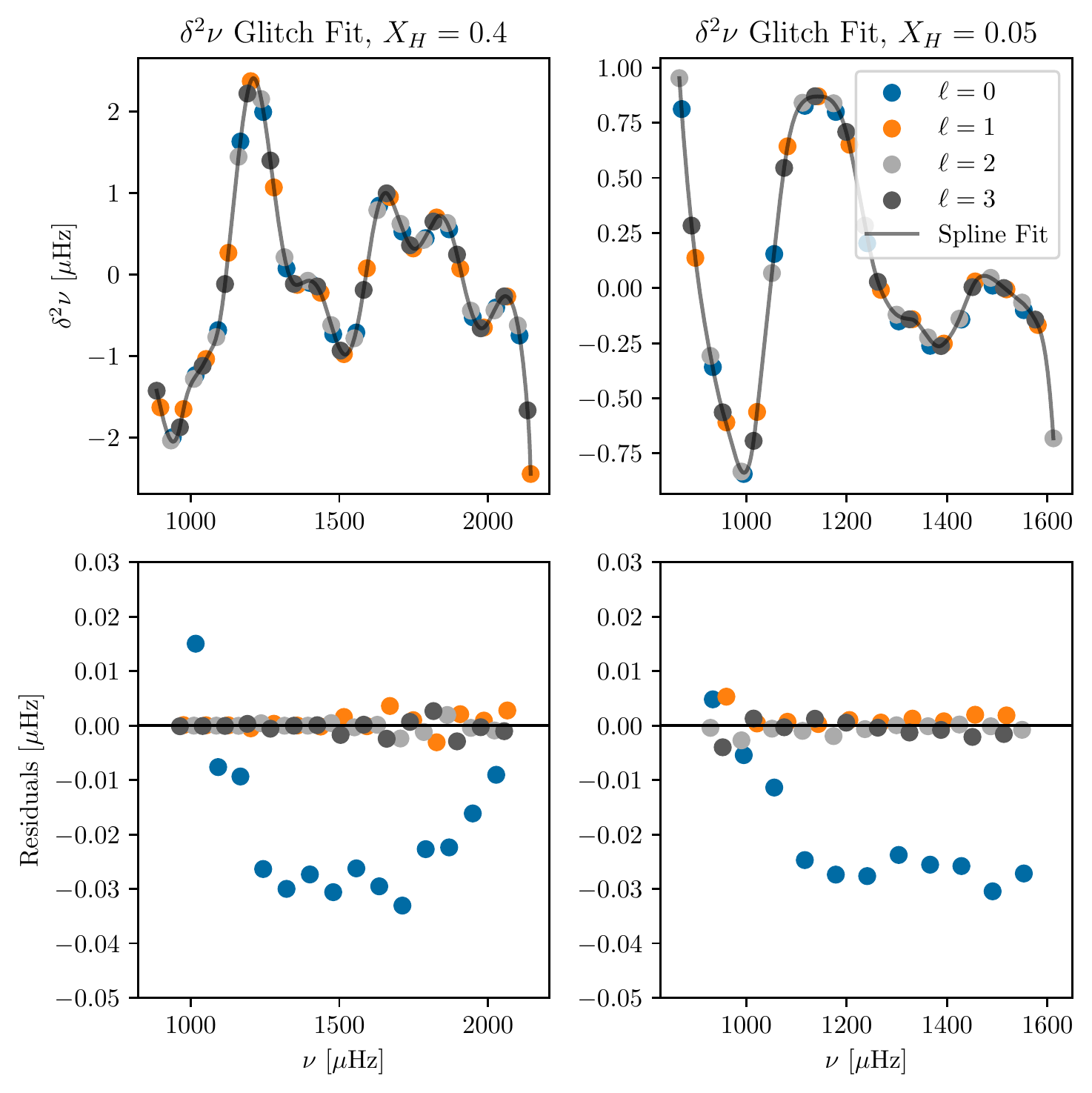}
    \caption{\textbf{Top Panels}: Plot of the second differences of the oscillation mode frequencies as a function of mode frequency for the same two 1.4 $M_{\odot}$ models as in \cref{fig:propagation}. The frequency ranges between 800 $\mu$Hz up to the acoustic cutoff frequency of the model ($\nu_{\text{ac}} = \frac{1}{2 \pi}\frac{c_{s} g \rho }{P} $) where the sound speed ($c_s$), density ($\rho$), and pressure ($P$) are taken at the model surface (outermost grid point). The second differences are taken for each set of $\ell = 0, 1, 2, \text{ or } 3$ modes with respect to the radial order $n_p$. The gray line shows a spline fit through the $\ell = 1, 2, \text{ and } 3$ modes. \textbf{Bottom Panels}: Corresponding residuals (second differences minus the spline fit) as a function of frequency.  }
    \label{fig:glitch}
\end{figure*}

Thus far, our discussion has taken place within the context of the commonly-used WKB approximation \citep[][]{Gough1993,Gough2007}. This is qualitatively suitable where the acoustic glitches are situated far enough away from the turning points of the mode cavity that the behavior of the wavefunctions there may be treated as approximately sinusoidal. However, at the turning points, the solutions are, instead, more accurately approximated by Airy functions, which relate to these sinusoidal solutions through the asymptotic expansion of the Airy functions at large argument, by way of Jeffreys connection formulae (i.e. the ``J" of JWKB). In turn, the use of Airy functions near the turning points is only justifiable when boundary conditions of the pulsation problem as a whole can be neglected. The resulting oscillatory variations induced into the mode frequencies from such analysis \cite[e.g.][]{Houdek2006} have, historically, been assumed to emerge even in existing theoretical studies of p-mode convective-core signatures \citep[e.g.][]{Monteiro1998,Mazumdar2001}. However, the near-core structural discontinuities in the mass range under consideration here do not possess these properties. Since these features, as well as the turning points themselves, are localized close to the core, the inner boundary condition may no longer be neglected. Since the glitches may not be inside the WKB-oscillatory region as is typically assumed, the wavefunctions likewise may not be well approximated as sinusoidal there. As such, we must pursue an alternative derivation of the frequency perturbations induced by these glitch features accounting for these properties, which may correspondingly yield qualitatively different behavior from the standard sinusoidal phenomenology.

\subsection{Analytic Development}

From local asymptotic theory, it is known that the scaled p-mode radial Lagrangian displacement wavefunctions, $\psi = \xi_r\sqrt{\rho r^2 c_s}$ near the center of the star may be approximated by linear combinations of Riccati-Bessel functions of degree $\ell$, with argument $\omega t$. These linear combinations are in turn well-described by using only the Bessel function of the first kind, with further position-dependent phases added to the argument. We refer the reader to \citet{calogero_novel_1963,babikov} for more details about this construction, and to \citet{Roxburgh2010,Ong2019} for more detailed discussion of the use of such phase functions in the context of p-modes. Here we use $s_\ell(x) = x j_\ell(x)$ to refer to the Riccati-Bessel functions of degree $\ell$ of the first kind, rather than the customary $S_\ell$, to avoid confusion with the Lamb frequency.

\edit1{We first} consider sharp variations in the Brunt–Väisälä frequency, relative to a smooth background: $N^2 = N^2_\text{smooth} + \delta N^2$, and wavefunctions that are unit-normalised under the usual inner product. \edit1{These sharp variations exist in our stellar models (see the $N$ profiles in figure \ref{fig:propagation}) and are collocated with enhancements to the acoustic potential $V$, which encapsulates all the relevant information for radial modes.} By inspection of the wave equations \citep[e.g.][and also \autoref{appendix2}]{Ong2020}, $\delta N^2$ induces deviations in the mode frequencies\edit1{, ceteris paribus,} as
\begin{equation}
\delta(\omega^2) \sim \int \xi_r^2 \cdot \delta N^2\ \mathrm d m\label{eq:N2kernel}
\end{equation}
compared to if only $N_\text{smooth}$ were present, to leading order in perturbation theory \cite[as also used in e.g.][]{Houdek2006,Houdek2007}. We first recount how the usual expression of these glitches, relating $\delta$-function features in the Brunt–Väisälä frequency, $\delta N^2 \sim \delta(r - r_0)$, to sinusoidal perturbations to the mode frequencies, may be recovered from this description. Near the outer boundary, $t = T$, the glitch signature of such a $\delta$-function feature may be computed with an approximate expression for the outer phase function of the form
\begin{equation}
\begin{aligned}
\delta (\omega^2) \sim \omega \delta\omega &\sim \int \mathrm d m\ \xi_r^2\ \delta(r - r_0) \sim \int \mathrm d t\ \psi^2(t)\ \delta(t - t_0) \\
&\sim s_\ell^2\left(\omega t_0 - \alpha_\ell\left(\omega, T - t_0\right) + \pi \left( n_p + \frac{\ell}{2}\right) - \omega T\right),
\end{aligned}
\end{equation}
where $r_0$ and $t_0$ are the physical and acoustic radii of the localized feature, $n_p$ is the radial order of the mode, and $\alpha_\ell$ is the phase function induced by the outer boundary condition. Far away from the center, the star may be well-approximated as being plane-parallel-stratified, and so the outer phase functions $\alpha_\ell$ do not materially depend on $\ell$ at low degree \citep[e.g.][]{Roxburgh2016}. The usual expression for acoustic glitches --- i.e. $\delta \omega \sim \sin \left[ 2 \omega (T - t_0) + \phi \right]$ for all $\ell$, up to some frequency-dependent amplitude function --- is then recovered upon introducing the asymptotic expansion of Riccati-Bessel functions as sinusoids at large argument: $s_l(x) \sim \sin\left(x - \pi{\ell \over 2}\right) + \mathcal{O}(1/x)$.

However, as we have described above, this usual derivation does not apply to these core acoustic glitches. Rather, since the glitches we consider are localized near the center of the star, we must instead make use of the converse expansion of Riccati-Bessel functions as power laws at small argument \citep[see][and \autoref{appendix1}]{arfken_weber_2005} :
\begin{equation}
s_l(x) \sim \frac{x^{\ell + 1}}{ (2 \ell + 1)!!},\text{ where } |x| \ll \sqrt{\ell + \frac{3}{2}},
\end{equation}
with the double exclamation marks denoting the semifactorial. Accordingly, the frequency perturbation induced by such near-core features takes the form 
\begin{equation} \label{eq:freq_perturbation}
    \delta \omega \sim \left([\omega t_0 - \delta_\ell(\omega, t_0)]^{\ell + 1} \over (2 \ell + 1)!! \right)^2,
\end{equation}
where $\delta_\ell(\omega, t)$  is an inner phase function induced by the inner boundary condition, satisfying $\delta_\ell(\omega, t)\to0$ as $t \to 0$ under regular boundary conditions at the center \citep[cf.][although we note that by using Riccati-Bessel functions here rather than sinusoids as in those works, we absorb the phase lag of $\pi \ell/2$ shown there --- cf. \autoref{appendix1}]{Roxburgh2010,Roxburgh2016}. This quantity can be seen to depend on $\ell$. Qualitatively, this implies that any frequency perturbation induced by a near-center feature must (1) diverge gradually with increasing frequency (as opposed to being sinusoidal, like outer glitches), and (2) possess an amplitude which decreases rapidly with increasing $\ell$ (as opposed to the $\ell$-independent behavior of outer glitches). In particular, since the semifactorial suppression with increasing $\ell$ is so steep, this effectively produces an offset of the radial-mode frequencies relative to all other $\ell$.

The variations to the Brunt–Väisälä frequency profile by themselves do not account completely for all structural variations at the boundary. For example, there are also variations to the sound speed profile at the convective core boundary, \citep[\edit1{seismic properties of which have been studied by, e.g.,}][]{Mazumdar2006, CunhaMetcalfe2007, Cunha2011} which could dominate the glitch signature for radial modes. In this work, we restrict our analysis to only a deviation in the Brunt–Väisälä frequency profile as we are interested in the qualitative properties of the near-core glitch signatures, namely their apparent \edit2{non-oscillatory nature and strong dependence on angular degree, $\ell$.} As demonstrated in \autoref{fig:propagation}, the sharp Brunt–Väisälä frequency features are collocated with enhancements to the acoustic potential, $V$, which also carries information about sound speed discontinuities.

\edit2{Similar analysis to the one done for the Brunt–Väisälä frequency, applied to the sound speed or other structural properties, will yield the same strong-degree-dependence behavior in the frequency perturbations. More precise statements concerning the exact frequency dependence and amplitudes of the mode frequency differences would require an analysis similar to \citet{CunhaMetcalfe2007}, incorporating structurally self-consistent perturbations to the relevant acoustic potentials. Perturbations to different quantities may yield different power-law indices in the frequency that may differ from that attributed to the Brunt–Väisälä frequency (if derivatives or integrals of the wavefunctions enter into the analogous kernel expressions to \cref{eq:N2kernel}), while those caused by different features will have the arguments of their power laws be evaluated at different acoustic depths. Thus, the overall frequency perturbation that we would expect from these near-core features will take the form of a sum of various power-law terms. However, we note that the argument and overall amplitude of any one of these power-law terms are in effect entirely degenerate. Thus, an unprivileged observer, given a combination of power law components resulting from near-core perturbations to the stellar structure, will find it mathematically impossible to distinguish the inner glitch depth from its amplitude.}

\subsection{Empirical Diagnostics}

\edit2{In the absence of a more quantitative description of the frequency dependence of the near-core glitch signatures,} we can illustrate \edit1{the qualitative properties of these} \edit2{near-core} glitch signatures \edit2{by computing} the mode frequencies for each stellar model along our 1.2 $M_{\odot}$, 1.3 $M_{\odot}$, 1.4 $M_{\odot}$, and 1.5 $M_{\odot}$ tracks using the stellar oscillation code GYRE \citep[version 6.0, ][]{Townsend2013}. We calculate the radial ($\ell = 0$) as well as non-radial ($\ell = 1, 2, \text{ and } 3$) mode frequencies in a \edit1{wide frequency range, from a lower bound of $\Delta \nu$ up to $2 \nu_{\text{max}}$.} We use scaling relations to approximate the global asteroseismic parameters of our stellar models based on the models' mass, radius, and temperature \citep[see][]{Kjeldsen1995}, setting $\nu_{\text{max}} = \nu_{\text{max}, \odot} M / (R^2 \sqrt{T_\mathrm{eff}})$ and $\Delta \nu = \Delta \nu_{\odot} \sqrt{M/R^3}$ where $M$, $R$, and $T_\mathrm{eff}$ are in solar units.


In order to enhance the visibility of these glitch signatures (oscillatory or otherwise), we take the second differences of the mode frequencies with respect to the modes' radial order $n_p$ \citep[$\delta^2{\nu_{n, \ell}}$, see][]{Gough1990, Basu1994, Basu2004, Mazumdar2005, Verma2014} given by, 
\begin{equation}
    \delta^2{\nu_{n, \ell}} = \nu_{n-1, \ell} - 2 \nu_{n, l} + \nu_{n+1, \ell}.
\end{equation}

For illustration, we show these second differences in the top panels of \cref{fig:glitch} for the same two 1.4 $M_{\odot}$ models as in \cref{fig:propagation} (before and after the discontinuous jump in the position of the p-mode inner turning point). As discussed previously, these can be seen to be dominated, for all $\ell$ shown, by the oscillatory variability of the outer ionization-zone/convective-boundary glitches, which should affect all $\ell$ equally, at least at these low degrees. We thus fit this overall oscillatory signal in the second differences using a cubic spline, incorporating only second differences of the $\ell = 1, 2, \text{ and } 3$ modes, as shown in the top panels of \cref{fig:glitch} as gray lines. We show the residuals to this fit in the bottom panels of \cref{fig:glitch}. The residuals are such that the $\ell = 0$ mode are clearly systematically offset from the non-radial mode frequencies. Thus, our analytic prediction for the amplitude of the glitch signature decreasing with increasing angular degree (Eq. \ref{eq:freq_perturbation}) is bourne out numerically. We note that this systematic dependence of the residuals on $\ell$ stays consistent between different choices for how the outer glitches are detrended (e.g. fitting a high-order polynomial instead of a spline, or also including $\ell = 0$ modes in the fit). Moreover, no obvious qualitative difference between the two stellar models can be seen, despite their being in different JWKB regimes (as earlier described). 

\begin{figure*}[htbp]
    \centering
    \includegraphics[width = 0.9\textwidth]{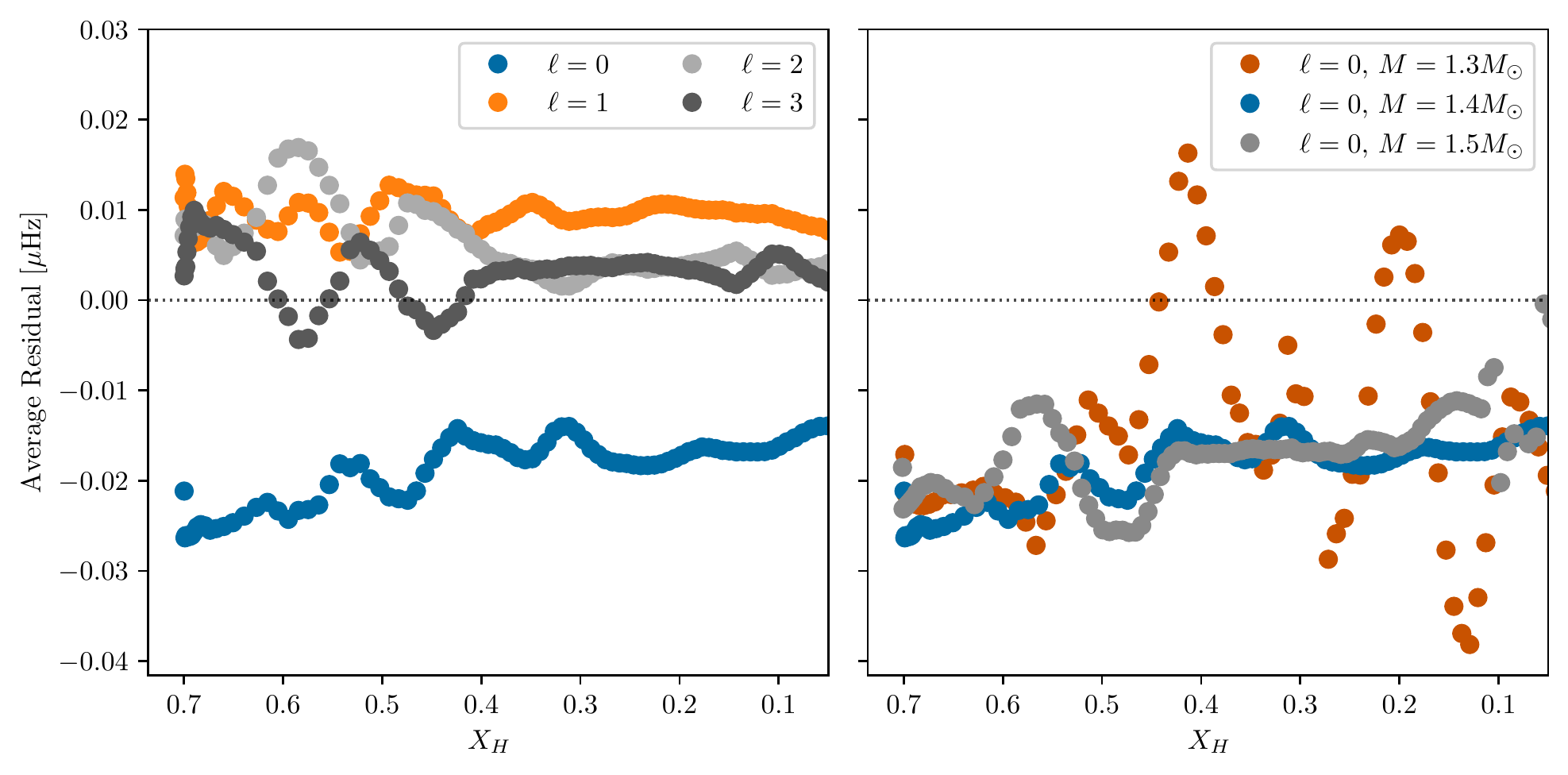}
    \caption{ \textbf{Left Panel:} The evolution of the 1.4 $M_{\odot}$ model track's average second-frequency-difference residuals left over after subtracting a high-order polynomial fit from the $\ell = $ 0, 1, 2, and 3 values of $\delta^2 \nu$. The evolution is shown as a function of the central Hydrogen fraction, $X_H$. \textbf{Right Panel:} The evolution of the $\ell = 0$ values of average $\delta^2 \nu$ residuals as a function of $X_H$ for our 1.3, 1.4, and 1.5 $M_{\odot}$ tracks.  \edit1{Each curved is smoothed with a boxcar kernel of size 5 points.} }
    \label{fig:residual_evolution}
\end{figure*}

To investigate the amplitude of the near-core glitch over the course of the main sequence evolution of our models, we computed the average second-frequency-difference residuals for each $\ell$ after subtracting a high-order polynomial fitted against $\delta^2\nu$ for $\ell = 0, 1, 2, 3$. For our 1.4 $M_{\odot}$ model track, we plot this average residual as a function of center Hydrogen fraction for each value of $\ell$ in the left panel of \cref{fig:residual_evolution}. Overall, the amplitude of the $\ell=0$ residuals is much larger than for the non-radial orders, agreeing with the analytic prediction that the amplitude of the frequency differences caused by the near-core glitch will decrease with increasing $\ell$ (\cref{eq:freq_perturbation}). The right panel of \cref{fig:residual_evolution} shows the evolution of the average $\ell = 0$ residuals for our 1.3, 1.4, and 1.5 $M_{\odot}$ model tracks. These appear very similar for the 1.4 and 1.5 $M_{\odot}$ tracks, and remain approximately constant over the course of their main sequence evolution.

\edit2{Our procedure for isolating the near-core glitch's affect on the second-difference residuals results in the radial-mode residuals containing, but not necessarily completely isolating, the near-core glitch signal. For example,} for much of the main sequence, the average $\ell = 0$, $\delta^2 \nu$ residual amplitudes for the 1.3 $M_{\odot}$ track are overall smaller when compared with the residual amplitudes for the 1.4 and 1.5 $M_{\odot}$ tracks, in keeping with the smaller size of the 1.3 $M_{\odot}$ models' convective cores. While \edit2{the 1.3 $M_{\odot}$ track residuals} may be seen to vary much more significantly after passing $X_H \approx 0.4$, this is not a feature of the near-core glitch, but rather a property of the outer glitches \edit2{contaminating the second-difference residual signal}. In particular, the convective envelope boundary of the 1.3 $M_{\odot}$ models is much deeper (in relative acoustic depth) compared with those of the 1.4 and 1.5 $M_{\odot}$ models. At around $X_H \approx 0.4$, the acoustic depth of the 1.3 $M_{\odot}$ model's convective envelope boundary increases past $\tau = T/2$ (where $T$ is the acoustic radius of the star). Interior to this, the glitch modulations \edit1{affect} not only the degree-independent outer phase function $\alpha_\ell$, but also \edit1{the} degree-dependent inner phase functions $\delta_\ell$ \cite[cf. Fig. 5 of][]{Roxburgh2010}; thus, the outer glitch may itself no longer be simply described as a function of frequency alone. As such, in this regime, the radial-mode residuals from such a fit will also contain contributions originating from the outer glitch, and no longer serve to describe the near-core glitch well. \edit2{Thus, we cannot guarantee that this method necessarily uniquely isolates the near-core glitch signal.}




\edit1{\citet{CunhaMetcalfe2007} have previously proposed an alternative means of eliminating the outer phase function by way of scaled separation ratios:}
\begin{equation} 
    \label{eq:diagnostic}
    \frac{D_{0, 2}}{\Delta\nu_{\text{n-1}, 1}} - \frac{D_{1, 3}}{\Delta \nu_{n, 0}},
\end{equation}
where
\begin{equation}
    D_{\ell, \ell+2} \equiv \frac{\nu_{\text{n},\ell} - \nu_{\text{n}-1, \ell + 2} }{4 \ell + 6}
\end{equation}
and
\begin{equation}
    \Delta \nu_{\text{n}, \ell} \equiv \nu_{\text{n}+1, \ell} - \nu_{\text{n}, \ell}.
\end{equation}
The un-scaled separation ratios ($r_{\ell, \ell+2} = d_{\ell,\ell + 2}/\Delta \nu_{\text{n}, \ell}$), considered as a function of frequency, were shown by \citet{Roxburgh2005} to be well-approximated by differences of the interior phase functions, $\delta_{\ell+2}(\nu) - \delta_{\ell}(\nu)$. The difference of the scaled ratios in Eq.~\ref{eq:diagnostic}, which is the diagnostic of \citet{CunhaMetcalfe2007}, is thus equivalent to taking a linear combination of the inner phase functions $\delta_0, \delta_1, \delta_2, \delta_3$, evaluated at some notional inner matching point, which is usually left underspecified. By contrast, \edit2{in} separating the acoustic depth from the inner phase function, \edit2{we evaluate the inner phase function} at the acoustic depth of the inner glitch. \edit2{A further, subtle} difference between these inner phase functions $\delta_\ell$ appearing here, and in our expressions, is that those we consider above are of the ``smooth" structure: they are therefore completely uninformative regarding the inner glitches, with all information about them contained instead in the $\omega t_0$ term, and the implicit constant of proportionality. By contrast, the $\delta_\ell$ of the procedure of \cite{CunhaMetcalfe2007} are associated with the actual mode frequencies, including the near-core glitches. \edit2{These differences render these two diagnostics not immediately quantitatively commensurate to each other, and deriving an explicit relationship between their diagnostic and ours lies beyond the scope of this work; at best, we will be able to perform only a qualitative comparison.}

\begin{figure*}[htbp]
    \centering
    \includegraphics[width = 0.45\textwidth]{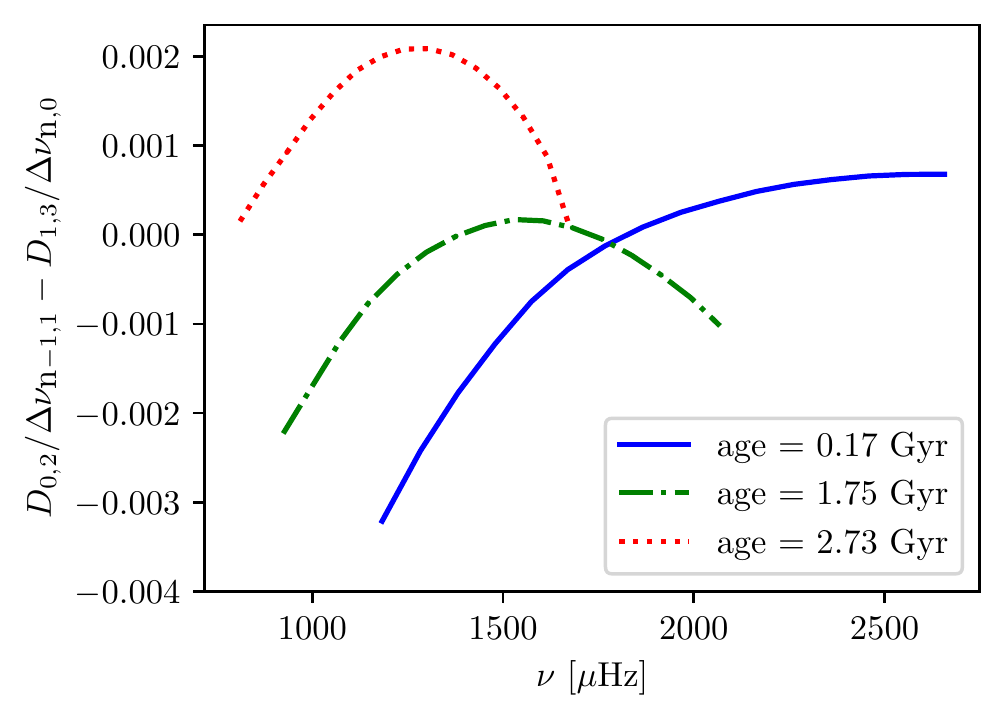}
    \includegraphics[width = 0.435\textwidth]{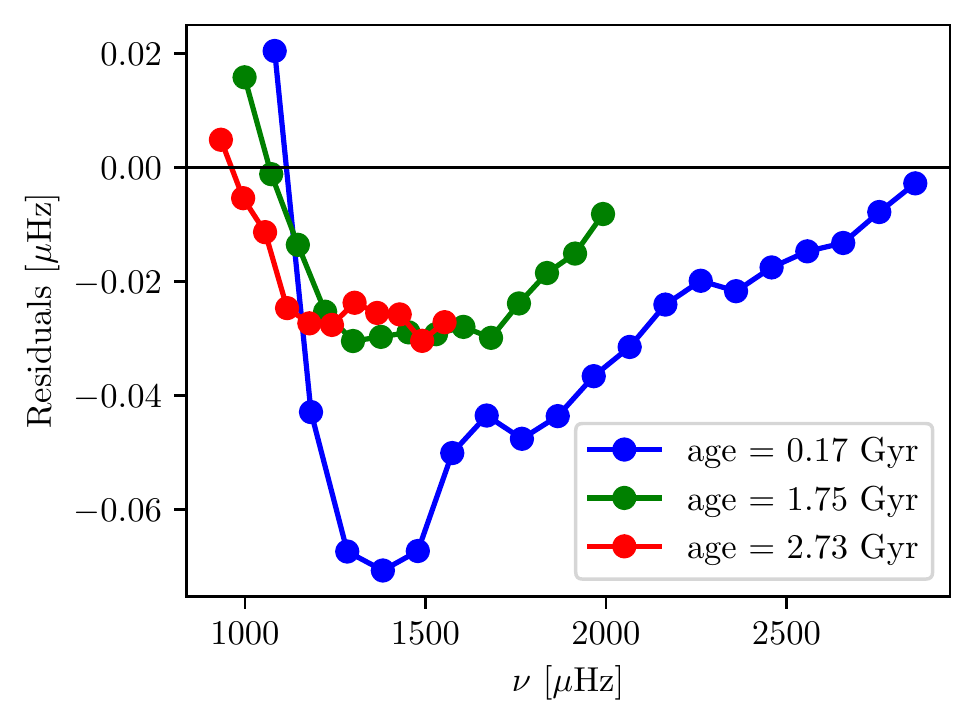}
    \caption{\textbf{Left Panel}: Frequency differences diagnostic defined by equation \ref{eq:diagnostic} as a function of frequency for three $1.4 M_{\odot}$ models at different ages. The model represented with the red dotted curve is the same as the model shown in the right panels of figure \ref{fig:glitch}. The model modes considered in this figure range in radial order from $n=8$ to $n=28$.  \textbf{Right Panel}: The second difference ($\delta^2 \nu$) residuals obtained after subtracting the outer glitch signature from just the radial mode values of $\delta^2 \nu$, as a function of frequency. The curves are shown for the same three models shown in the left panel.}
    \label{fig:Cunha3}
\end{figure*}

Plotting the diagnostic from Eq.~\ref{eq:diagnostic} as a function of frequency (figure \ref{fig:Cunha3}a) for three $1.4 M_{\odot}$ stellar models of different ages shows that this diagnostic, calculated for our model frequencies, shows similar properties to those shown \citet{CunhaMetcalfe2007}, despite differences in the modelling physics and global stellar properties. Comparing the residuals of our outer glitch subtracting procedure (from figure \ref{fig:glitch}) for the same three models (shown in figure \ref{fig:Cunha3}b) with the diagnostic curves in figure \ref{fig:Cunha3}a, shows the curvature is reversed between the two methods of displaying the inner, near-core glitch signal. However, both methods show that subtracting the outer glitches is necessary to reveal the small-amplitude seismic signatures of the convective cores. In addition, \edit2{both panels of} \ref{fig:Cunha3} show that both methods of isolating the seismic signal of the near-core glitch on the radial modes reveals that these signals are age-dependent, meaning their glitch signature shape changes as a main sequence star evolves on the main sequence.

\edit2{Practically speaking , an operational} difference between \edit2{the diagnostic in} Eq.~\ref{eq:diagnostic} and our method of subtracting a spline fit from the non-radial modes detailed in section \ref{sec:nonwkb} is that octopole ($\ell = 3$) modes are not explicitly required by our construction. \edit2{In principle, our proposed methodology accommodates data sets containing both fewer, and more, degrees than $\ell \in \{0,1,2,3\}$}. However, retrieving the asteroseismic signal of the convective cores would still be difficult unless many non-radial mode frequencies are available.

\section{Discussion and Conclusion} \label{sec:discussion}

Our analysis shows two distinct WKB regimes for the solar-like oscillators in question. Throughout the first regime, before the discontinuous increase in $R_{\ell=1}$, the realm that non-radial p-mode oscillations probe includes the near-core regions of the star, which is of particular importance for studies of \edit2{convective boundary} mixing processes like convective overshoot. During the second regime, after the sharp increase in $R_{\ell=1}$, the near-core layers around the stellar core are no longer accessible to non-radial p-mode oscillations with frequencies near $\nu_{\text{max}}$. This suggested that the usefulness of these modes to study core processes in main sequence stars would depend on the exact evolutionary history of that particular object, due to the sharp boundary between the two regimes. Acoustic glitch fitting is often used to determine the locations of particular stellar layers of interest, such as the boundaries of convection zones and the locations of ionization zones. In the first regime we discuss, where the p-mode outer turning point is exterior to the boundary of the well-mixed core, acoustic glitch fitting would be used to study the different layers of the star down past the boundary of the well-mixed core. In this case, glitch signatures from the boundary of the convective core will impart perturbations to the (near-$\nu_{\text{max}}$) frequencies of both the radial and non-radial oscillation modes of the star. On the other hand, assuming the WKB approximation holds, after the discontinuous jump in $R_{\ell=1}$, the near-core glitch signature from the core convection zone boundary should be inaccessible to acoustic glitch analysis.

However, since the well-mixed convective core boundary exists sufficiently close to the center of the star such that the inner boundary condition can no longer be neglected, and $R_c$ may not be inside the WKB-oscillatory region in any case, the glitch signature pattern from the core boundary imprints instead an $\ell$-dependent signal onto the frequencies of the stellar oscillation modes. We demonstrate this behavior in \cref{fig:glitch} where after fitting out the dominant acoustic glitch signature from the $\ell = $ 0, 1, 2, and 3 modes; an additional glitch signature is visible in the radial-mode residuals in both cases, which in neither case appears oscillatory. \edit2{The results of this procedure can be seen to qualitatively resemble those obtained from other prior proposed diagnostics, such as that of \citet{CunhaMetcalfe2007} (\autoref{fig:Cunha3}).} 

\edit2{In contrast to the oscillatory and degree-independent nature of acoustic glitches in the outer parts of the star, we would therefore expect observationally to ultimately obtain, from isolating the inner glitch signal, some combination of non-oscillatory components, each exhibiting some power-law nature and strong angular degree dependence in their frequency perturbations. Crucially for the outer glitches, it is precisely their sinusoidal form which permits different components, localized at and attributed to different physical features, to be separately identified and characterized through their modulation frequency and amplitude \citep{Houdek2006, Monteiro1998, Mazumdar2001}. By contrast, since the amplitude and argument of a power law are mathematically indistinguishable, it is impossible to distinguish the amplitude of the inner glitch from its argument (i.e. location, via acoustic depth), let alone disentangle and assign interpretations to such linear combinations of them as we should expect to obtain in practice, without the imposition of further constraints from stellar modelling. Thus, our key qualitative result in this work --- that inner glitches hold to power-law parameterizations --- also indicates that any quantitative signal, however isolated observationally, will not be amenable to as easy interpretation as those derived from the outer glitches. Unfortunately, this means that any insights into the nature of near-core convective boundary mixing must necessarily derive from explicit reference to numerical models of stellar structure, unlike the model-independent diagnostic quantities returned from the outer glitches.}

We note that the aforementioned \edit1{near-core glitch properties we discuss in this work} are only applicable to stars which are massive enough to host convective cores, but with low enough masses to have significant convective envelopes which drive p-mode oscillations. \edit1{Therefore, future searches for seismic signatures of main-sequence small convective cores will need to limit their consideration to stars within a narrow mass range with good asteroseismic data.}

Stellar structures like those discussed in this paper, characterized by steep, localized variations in structure near their cores, are not just present in intermediate mass main sequence stars. As low and intermediate mass stars run out of hydrogen in their cores and begin to evolve across the subgiant branch and up the red giant branch, their convective envelopes expand while their cores contract \citep[cf.][]{2017A&ARv..25....1H}. At this stage of evolution, the interior boundary of the convective envelope reaches far into the core of the giant star, depending on the amount of envelope overshooting \citep[see §4, Figure 4 of][]{Lindsay2022}. The steep variation in sound speed at the interior boundary of the envelope convection zone will induce a glitch component to the mode frequencies of the giant star and, since the location of the glitch would be near the core in this case, an $\ell$-dependent signature similar to that described in \autoref{sec:nonwkb} will be likewise present in the notional p-modes of these giants. Since the observable modes in red giant stars are in practice modes of mixed g- and p-like character, disentangling the deep envelope convection zone glitch signature from the overall mixed mode pattern of observed red giant oscillation modes would be challenging, but rewarding. Such constraints on the location of the envelope convective boundaries, which would define the correct amount of envelope overshooting, which should be incorporated into evolved star stellar models. These would be complementary to other constraints from ``buoyancy" glitches, derived from the g-mode cavity \citep[e.g.][]{cunha_analytical_2019,vrard_evidence_2022}. At the same time, the considerations we outline here may be required to interpret such buoyancy glitches: should they be localized near the g-mode turning points (as we describe in \citealt{Lindsay2022}), the relevant wavefunctions should be described with Airy functions of the first kind, \edit2{as also used in \citet{CunhaMetcalfe2007}}, which would yield different behavior from the sinusoids considered in \cite{cunha_analytical_2019}.

\begin{acknowledgements}
CJL acknowledges support from a Gruber Science Fellowship. JO acknowledges support from NASA through the NASA Hubble Fellowship grant HST-HF2-51517.001 awarded by the Space Telescope Science Institute, which is operated by the Association of Universities for Research in Astronomy, Incorporated, under NASA contract NAS5-26555. SB acknowledges NSF grant AST-2205026.  We also thank Dan Hey, Marc Hon, and the anonymous referee for their useful and constructive discussion, as well as the Yale Center for Research Computing for guidance and use of the research computing infrastructure.

\software{
MESA \citep{Paxton2011,Paxton2013,Paxton2015,Paxton2018,Paxton2019}, GYRE \citep{Townsend2013}},
SciPy \citep{scipy}, Pandas \citep{pandas}

\edit1{The MESA and GYRE inlists we used to generate our models and frequencies, as well as the resulting MESA models and frequencies, are archived on Zenodo and can be downloaded at 
 \dataset[https://doi.org/10.5281/zenodo.7705648.]{https://doi.org/10.5281/zenodo.7705648}}
\end{acknowledgements}

\appendix 
\section{Spherical Bessel Functions}
\label{appendix1}
The usual expression for the mode frequency shifts resulting from an acoustic glitch has the form $\delta \omega \sim \sin[2 \omega (T - t_0) + \phi]$. However, in the case where the acoustic glitch is located sufficiently close to the real (coordinate) singularity at the center of the star, the usual expression for acoustic glitches does not apply. In section \ref{sec:nonwkb}, we made use of the converse expansion of Riccati-Bessel functions as power laws at small argument. Here we expand on this discussion of spherical Bessel functions, drawing extensively from §11.7 of \citet{arfken_weber_2005}. 

Upon separation of variables of the wave equation, the radial wavefunction $R$ satisfies an ordinary differential equation that is well-approximated near the center of the star by an expression of the form
\begin{equation}
    x^2 \frac{d^2 R}{dx^2} + 2x \frac{dR}{dx} \sim - \big[x^2 - \ell(\ell+1)\big] R,
\end{equation}
where $\ell(\ell+1)$ is the separation constant from the angular components ($\ell$ is a non-negative integer) and the dimensionless argument $x \sim k_r r$ enters from the coordinate transformation required to put the original Helmholtz equation into this form. If one makes the substitution $R(x) = \frac{Z(x)}{\sqrt{x}}$, the radial equation becomes
\begin{equation}
    x^2 \frac{d^2 Z}{dx^2} + x \frac{d Z}{dx} + \left[x^2 - \left(\ell + \frac{1}{2}\right)^2 \right] Z = 0,
\end{equation}
which is Bessel's equation with $Z$ being a Bessel function of order $\ell + \frac{1}{2}$. Defining 
\begin{equation}
    j_\ell(x) = \frac{s_{\ell}(x)}{x} = \sqrt{\frac{\pi}{2x}} J_{\ell + 1/2}(x)
\end{equation}
and expressing $J_{\ell}$ as a series \citep[c.f. §11.1 of][]{arfken_weber_2005}:
\begin{equation}
    J_{\ell + 1/2}(x) = \sum_{s = 0}^{\infty} \frac{(-1)^s}{s!(s + \ell + \frac{1}{2})} \bigg(\frac{x}{2}\bigg)^{2s + \ell + \frac{1}{2}} = \frac{x^\ell}{2^\ell \ell!} - \frac{x^{\ell + 2}}{2^{\ell + 2} (\ell + 1)! }+ ...
\end{equation}
we apply the Legendre duplication formula, $z! \left(z + \frac{1}{2}\right)! = 2 ^{- 2z -1} \sqrt{\pi} (2z + 1)!$, to each term. Thus, we have a series form for $j_{\ell}(x)$, 

\begin{equation}
    \begin{aligned}
    j_{\ell}(x) & = \sqrt{\frac{\pi}{2x}} \sum_{s=0}^{\infty} \frac{(-1)^s 2^{2s + 2\ell + 1} (s + \ell)!}{\sqrt{\pi}(2s + 2\ell + 1)! s!} \bigg( \frac{x}{2}\bigg)^{2s + \ell + \frac{1}{2}} \\
     & = 2^{\ell} x^{\ell} \sum_{s=0}^{\infty} \frac{(-1)^{s}}{s! (2s + 2 \ell + 1)!} x^{2s}.
    \end{aligned}
\end{equation}

Now in the limit of small argument, where $x \ll 2\sqrt{\frac{(2\ell + 2)(2\ell + 3)}{(\ell+1)}}$, we have 
\begin{equation}
    j_{\ell}(x) \approx \frac{2^\ell \ell!}{(2 \ell + 1)!} x^{\ell} = \frac{x^\ell}{(2 \ell + 1)!!}.
\end{equation}
The expressions in \autoref{sec:nonwkb} are then recovered with argument $x = \omega t_0 - \delta_\ell(\omega, t_0)$.

\begin{figure}
    \centering
    \includegraphics{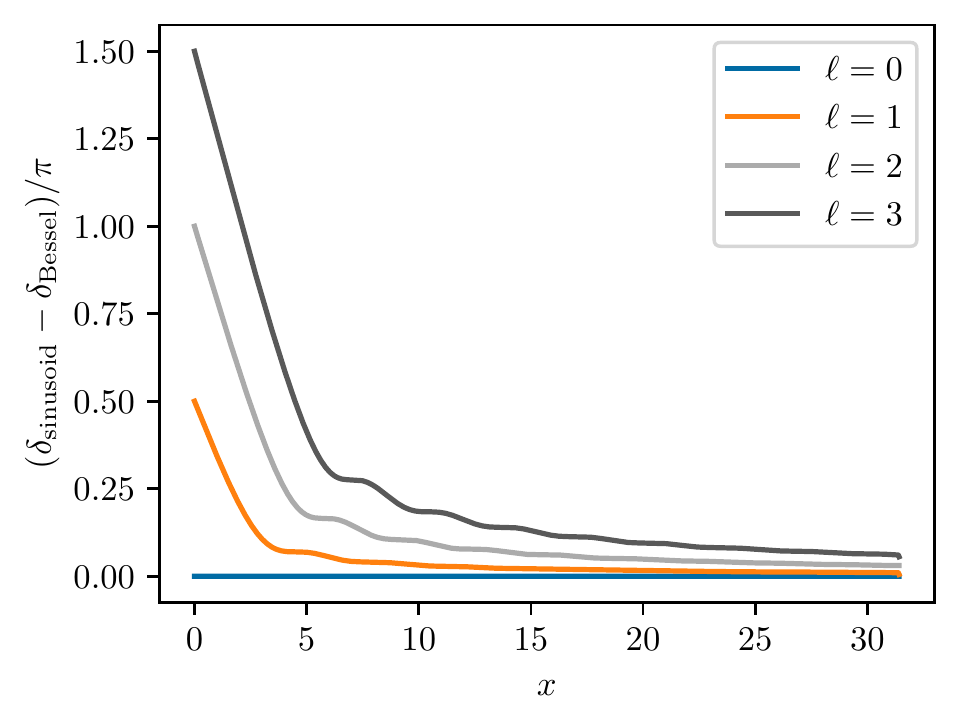}
    \caption{The inner phase function $\delta$ required to approximate a Riccati-Bessel function using a sinusoid.}
    \label{fig:delta}
\end{figure}

Our use of the Riccati-Bessel functions here requires also that our inner boundary condition for $\delta$ differs from that of \citet{Roxburgh2010,Roxburgh2016}, who use a sinusoidal approximation, such that the phase function required to approximate a wavefunction $\psi$ with a sinusoid as $A \sin \left(\omega t + \delta\right)$ can be found as $\delta \sim \mathrm{arctan}\left(\psi \left/ {\mathrm d \psi \over \mathrm d (\omega t)}\right.\right)  - \omega t$. For illustration, we show this in \cref{fig:delta} for Riccati-Bessel functions $s_\ell$ of various degree. This is known to yield an offset of $\ell \pi / 2$ in the argument as $t \to 0$, which is exactly equal to the inner boundary condition of \citet{Roxburgh2010,Roxburgh2016}. Thus, $\delta_\ell \to 0$ as $t \to 0$ for all $\ell$ in our description, for consistency with these works further into the stellar interior.

\section{Deviations to the mode frequencies}
\label{appendix2}
\edit1{
The displacement eigenfunctions $\bm{\xi}$ of normal modes with angular frequency $\omega$ satisfy the constraint
\begin{equation}
    -\rho \omega^2 \bm{\xi} = -\nabla P' + \mathbf{g} \rho' + \rho \nabla \Phi',\label{eq:xi}
\end{equation}
where $P'$, $\rho'$, and $\Phi'$ are the accompanying eigenfunctions in the pressure, density, and gravitational potential perturbations for that mode. These other eigenfunctions may be eliminated with the use of other physical constraint equations to yield an operator eigenvalue equation, customarily written in the manifestly Hermitian form
\begin{equation}
\begin{aligned}
    -\rho \omega^2 \bm{\xi} &= \nabla \left(\bm{\xi} \cdot \nabla P + c_s^2 \rho \nabla \cdot \bm{\xi}\right) - \mathbf{g} \nabla \cdot (\rho \bm{\xi}) - \rho G \nabla\left(\int \mathrm d^3 x' {\nabla \cdot(\rho \bm{\xi}) \over |x - x'|}\right)\\
    &\equiv \rho \hat{\mathcal{L}} \bm{\xi}. \label{eq:hermitian1}
\end{aligned}
\end{equation}
Small (and necessarily Hermitian) perturbations to the wave operator, of the form $\hat{\mathcal{L}} \mapsto \hat{\mathcal{L}} + \lambda\hat{\mathcal{V}}$, then yield perturbations to the mode frequencies as
\begin{equation}
    \delta (-\omega_i^2) \sim \lambda V_{ii} + \lambda^2 \sum_{j \ne i} {|V_{ij}|^2 \over \omega^2_{0, i} - \omega^2_{0, j}} + \mathcal{O}(\lambda^3), \label{eq:expansion}
\end{equation}
where $V_{ij} = \int \mathrm d^3 r\  \rho \bm{\xi}^*_i \cdot \hat{\mathcal{V}} \bm{\xi}_j$ are the matrix elements of the perturbing operator, and the Lagrangian displacement functions are assumed to be unit-normalized. For instance, $\hat{\mathcal{V}}$ may be considered to be the difference between the wave operators of two different stellar structures with identical global properties, such that the matrix elements may be expressed as integrals with respect to localized perturbations in the physical quantities of the stellar structure \cite[e.g. the inversion kernels of ][]{Kosovichev1999}. In discussions of acoustic glitches, however, one instead supposes that the wave operator $\hat{\mathcal{L}}$ may be notionally decomposed as $\hat{\mathcal{L}}_\text{smooth} + \hat{\mathcal{V}}_\text{sharp}$. The first term is, in the abstract, the wave operator associated with a ``smooth" stellar structure, such that (by assumption) its eigenfunctions are well-described by asymptotic approximations such as the JWKB construction, while the second term is associated with localized, sharp, variations in the stellar structure.
Since such a decomposition is at best notional, we are free to consider expressions for $\hat{\mathcal{V}}_\text{sharp}$ which might otherwise correspond to unphysical structural perturbations in the traditional sense. Moreover, by \cref{eq:expansion}, we may restrict our attention to the matrix elements of various operators, rather than the operators themselves. In particular, we note that a subset of the terms in \cref{eq:hermitian1}, which we shall use to define an operator $\hat{\mathcal{H}}$, have matrix elements
\begin{equation}
    H_{12} = \left<\bm{\xi}_1, {1 \over \rho} \left(\nabla \left(\bm{\xi}_2 \cdot \nabla P\right) - \mathbf{g} \nabla \cdot (\rho \bm{\xi}_2) \right) \right> = -\int \mathrm d^3 x\ \left[\bm{\xi}_1 \cdot \mathbf{g} \nabla \cdot (\rho \bm{\xi}_2) + \bm{\xi}_2 \cdot \mathbf{g} \nabla \cdot (\rho \bm{\xi}_1) - (\bm{\xi}_2 \cdot \mathbf{g})(\bm{\xi}_1 \cdot \nabla \rho) \right]
\end{equation}
that are Hermitian, by the divergence theorem (and since both $\mathbf{g}$ and $\nabla \rho$ point strictly radially). Focusing on the first two terms in particular, the constraints of adiabaticity $\rho' = {{P' \over c_s^2} + \rho(\mathbf{e}_r \cdot \bm{\xi}){N^2 \over g}}$, and of continuity $\rho' = -\nabla \cdot \left(\rho \bm{\xi}\right)$, allow us to rewrite this expression as
\begin{equation}
    \left<\bm{\xi}_1, \hat{\mathcal{H}}\bm{\xi}_2\right> = \int \rho\ \mathrm d^3 r \left(-2N^2 (\mathbf{e}_r \cdot \bm{\xi}_1)(\mathbf{e}_r \cdot \bm{\xi}_2)\right) + \text{\textit{(other Hermitian terms)}}.
\end{equation}
Accordingly, if we were to consider a notional, unphysical decomposition of $\hat{\mathcal{L}}$ as above, in which for $\hat{\mathcal{L}}_\text{smooth}$ only this Brunt-Vaisala frequency term were to be modified as $N^2 \mapsto N^2_\text{smooth} + \delta N^2$, the corresponding perturbation induced into the mode frequencies, ceteris paribus, would then go as
\begin{equation}
\delta(\omega^2) \sim \left<\bm{\xi}, \delta\hat{\mathcal{H}}\bm{\xi}\right>\sim \int \xi_r^2 \cdot \delta N^2\ \mathrm d m.
\end{equation}
More principled decompositions necessarily have a more complicated form. For instance, one might prefer to consider frequency differences arising from more traditional perturbations to the equilibrium $\rho$, $\Gamma_1$, $P$, etc. in a physically and structurally self-consistent fashion, for which the frequency differences arising from perturbations to specific quantities $q$ are associated with integral kernels of the form
\begin{equation}
    V_{ij} = \int (\delta q/q)\ \ K_q[\bm{\xi}_i, \bm{\xi}_j]\ \mathrm d m.
\end{equation}
By inspection of \cref{eq:hermitian1} these structure kernels, $K$, must necessarily be bilinear (and therefore quadratic, on the diagonal) in the wavefunctions of the modes corresponding to each matrix index, or potentially their (anti)derivatives with respect to radial position, as $\hat{\mathcal{V}}$ is permitted to be a general integro-differential operator. Since the asymptotic radial dependence of the (appropriately scaled) wavefunctions is as a power law as $r \to 0$, as we describe above, the overall signature of the near-core feature would then be a linear combination of components, each satisfying a power-law description of the kind we have provided. Our qualitative results are thus not substantially changed were a different quantity to be primarily responsible for producing the acoustic glitch (although each component may have an incremented or decremented power-law index, or different constant of proportionality).
}

\end{document}